\documentclass{article}
\usepackage{dcase2020_techrep,amsmath,graphicx,url,times,booktabs, tabularx, multirow}

\title{Sound Event Localization and Detection Using\\
Activity-Coupled Cartesian DOA Vector and RD3net}

\name{Kazuki Shimada, Naoya Takahashi, Shusuke Takahashi, Yuki Mitsufuji}
\address{Sony Corporation, Japan}

\begin{document}

\ninept
\maketitle

\begin{sloppy}

\begin{abstract}
% Sound event localization~(SEL) and sound event detection~(SED) are recognized as important tasks for human auditory cognition in real life and have been investigated intensively to date. To address these challenging tasks, we studied DCASE2020 task~3: Sound Event Localization and Detection (SELD). We propose a unified training framework that uses an activity-coupled Cartesian DOA vector~(ACCDOA) representation as a single target for both the SED and SEL tasks. ACCDOA assigns an audio event activity to the length of a corresponding Cartesian DOA vector. 
% The model using the ACCDOA representation is trained to minimize the distance between the estimated and the target coordinates. To efficiently estimate sound event locations and activities, we further propose RD3Net, which incorporates recurrent and convolution layers with dense skip connections and dilation. To generalize the network, we apply three data augmentation techniques: equalized mixture data augmentation~(EMDA), rotation of first-order Ambisonic~(FOA) singals, and multichannel extension of SpecAugment. Furthermore, we also use RD3Net for a two-stage system that solves two tasks independently and ensembles the model outputs. Our systems demonstrate a significant improvement over the baseline method.
Our systems submitted to the DCASE2020 task~3: Sound Event Localization and Detection (SELD) are described in this report. We consider two systems: a single-stage system that solve sound event localization~(SEL) and sound event detection~(SED) simultaneously, and a two-stage system that first handles the SED and SEL tasks individually and later combines those results. As the single-stage system, we propose a unified training framework that uses an activity-coupled Cartesian DOA vector~(ACCDOA) representation as a single target for both the SED and SEL tasks. To efficiently estimate sound event locations and activities, we further propose RD3Net, which incorporates recurrent and convolution layers with dense skip connections and dilation. To generalize the models, we apply three data augmentation techniques: equalized mixture data augmentation~(EMDA), rotation of first-order Ambisonic~(FOA) singals, and multichannel extension of SpecAugment. Our systems demonstrate a significant improvement over the baseline system.

\end{abstract}


\section{Introduction}
\label{sec:intro}
% Sound event localization and detection~(SELD) is the task of identifying both the direction of arrival~(DOA) and the type of sound.  A number of methods have been tackling this challenging problem by decomposing tasks into several subtasks: the estimation of the number of sources, DOA estimation, and sound event detection~(SED). Although this simplifies the SELD problem and therefore could improve the performance of each task, it also increases system complexity and computational cost. Here, we introduce a unified framework that uses an activity-coupled Cartesian DOA vector (ACCDOA) representation as a single target for the SED and sound event localization~(SEL) tasks. ACCDOA assigns an audio event activity to the length of a corresponding Cartesian DOA vector. The model using the ACCDOA representation is trained to minimize the distance between the estimated and the target coordinates.

Sound event localization and detection~(SELD) is the task of identifying both the direction of arrival~(DOA) and the type of sound. A number of methods have been tackling this challenging problem by decomposing tasks into several subtasks: the estimation of the number of sources, DOA estimation, and sound event detection~(SED). Although this simplifies the SELD problem and therefore could improve the performance of each task, it also increases system complexity and computational cost. Here, we consider both the single- and two-stage systems. The single-stage system solve both SED and SEL task simultaneously using an activity-coupled Cartesian DOA vector (ACCDOA) representation. ACCDOA assigns an audio event activity to the length of a corresponding Cartesian DOA vector. On the other hand, the two-stage system first handles the SED as a frame wise classification problem and then combines with the DOA estimation.

\section{System}
\label{sec:system}
In this section, we first give an overview of our two systems, namely, the ACCDOA (single-stage) system and the two-stage system. Then we explain the parts of our pipelines: the features, data augmentation, network architecture, and loss function.

\subsection{System overview}
\label{ssec:system}
A schematic flow of the ACCDOA system is shown in Fig.~\ref{fig:overview}. Two data augmentation techniques are applied to input signals prior to the feature extraction while one data augmentation technique exploiting multichannel information in the feature domain is performed after the feature extraction. Finally, the network outputs frame-wise ACCDOA vectors for 14 sound events. The magnitude of the vectors corresponds to the probability of each sound event activity while the direction of the vectors points toward the location of each source. The model using the ACCDOA representation is trained to minimize the distance between the estimated and the target coordinates.

\begin{figure}[t]
    \centering
    \centerline{\includegraphics[width=0.88\linewidth]{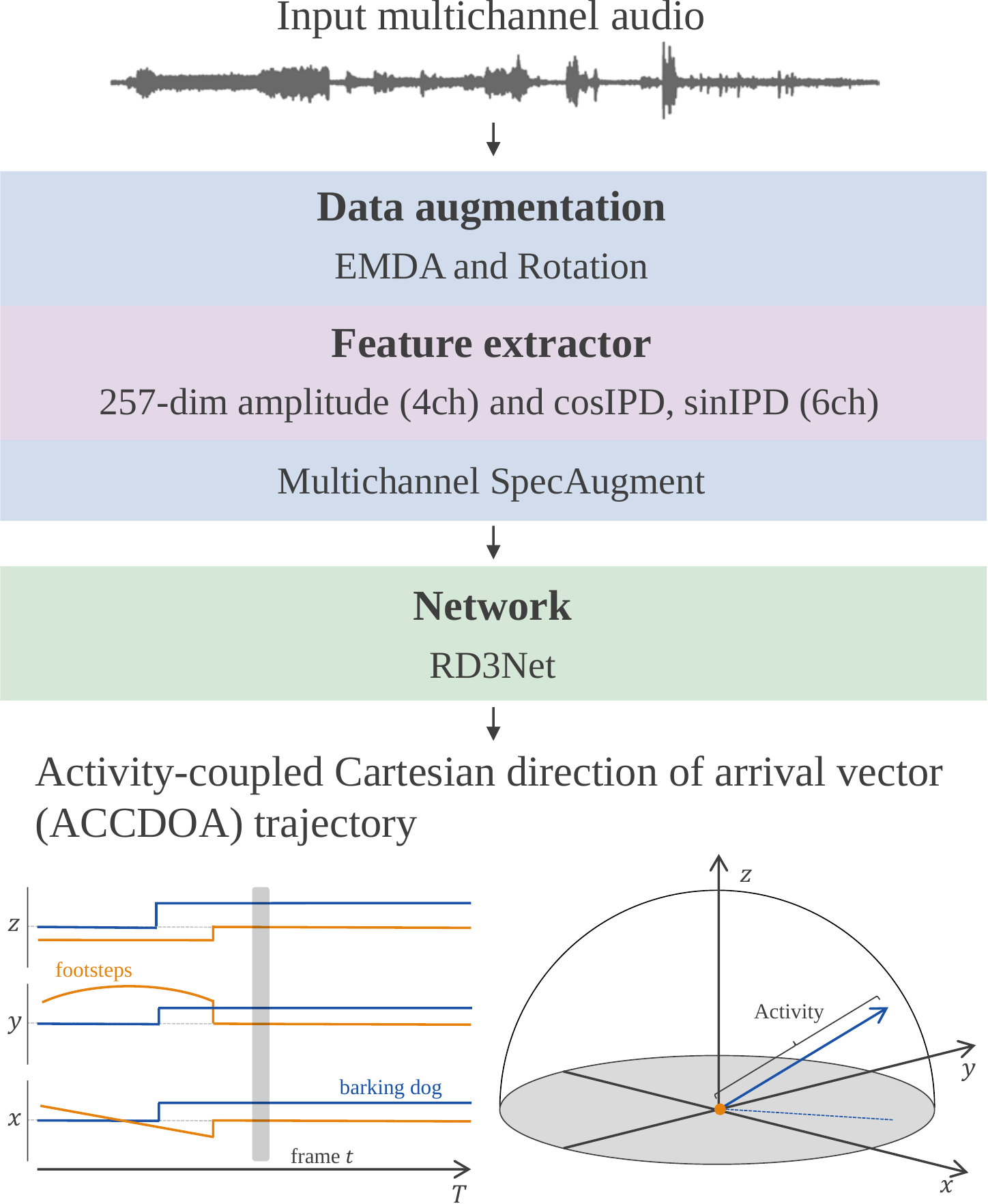}}
    \caption{Illustration of our ACCDOA system.}
    \label{fig:overview}
    \vspace{-2mm}
\end{figure}

% Apart from the ACCDOA system, we introduce another system inspired by the two-stage scheme proposed in~\cite{cao2019polyphonic}. 
The two-stage system is inspired by the work in~\cite{cao2019polyphonic}. 
This is characterized by three ideas: training only the SED branch, transferring a part of the network parameters from the SED branch to the DOA estimation branch, and training the DOA estimation branch. The data augmentation techniques, the features, and the network used in the system are exactly the same as the ones in the ACCDOA system.   

\subsection{Feature}
\label{ssec:feature}
Multichannel amplitude spectrograms and inter-channel phase differences~(IPDs) are used as frame-wise features. Here, ${\rm IPD}_{t, f, p, q} = \angle{x}_{t, f, p} - \angle{x}_{t, f, q}$ is computed from the short-time Fourier transform~(STFT) coefficients $x_{t, f ,p}$ and $x_{t, f, q}$, where $t, f, p,$ and $q$ denote the time frame, the frequency bin, the microphone channel $p$, and the channel $q$, respectively.
We fix $p=0$ to compute relative IPDs between all the other channels, $q$.
Since the input consists of four channel signals, we can extract four amplitude spectrograms and three IPDs.

\subsection{Data augmentation}
\label{ssec:augmentation}
To promote the generalization of the model, we exploit the following data augmentation techniques during the training.
\begin{itemize}
    \item \textbf{EMDA}: As described in \cite{Takahashi16,Takahashi2017AENet}, we apply the equalized mixture data augmentation~(EMDA) method where up to two audio events are mixed with random amplitudes, delays, and the modulation of frequency characteristics, i.e. equalization.
    \item \textbf{Rotation}: We also adopt the spatial augmentation method in~\cite{mazzon2019first}.  It rotates the training data represented in the first order Ambisonic~(FOA) format and allows us to increase the numbers of DOA labels without losing the physical relationships between steering vectors and observations. 
    % To avoid going out of range, we use only reflection for elevation, while for azimuth we use the combinations of reflection and rotation of 180 degrees.
    We consider eight rotation patterns for azimuth $\phi$ and elevation $\theta$: $(\phi', \theta') = (\phi, \theta), (-\phi, \theta), (\phi + \pi, \theta), (-\phi + \pi, \theta), (\phi, -\theta), (-\phi, -\theta), (\phi + \pi, -\theta), (-\phi + \pi, -\theta)$.
    \item \textbf{Multichannel SpecAugment}: We propose a multichannel version of SpecAugment in~\cite{park2019specaugment, zhang2019data}. In addition to the time-frequency hard masking schemes applied on amplitude spectrograms, we also extend it to the channel dimension. The target channel for the channel masking, $c_{0}$, is chosen from $[0, C)$ where $C$ denotes the number of microphone channels. For the IPD features, instead of multiplying a mask value by the original value, the original values are replaced with random values, where the values are sampled from a uniform distribution ranging from 0 to $2\pi$.
\end{itemize}

\subsection{Network architecture}
\label{ssec:network}
As the network architecture, we adopt the D3Net architecture~\cite{Takahashi20d3net}, which has shown the state-of-the-art performance in music source separation. The adaptation to the SELD problem includes three modifications. First, we omit dense blocks in the up-sampling path because high frame-rate prediction is not necessary for the SELD problem. Second, we replaced with GRU cells only in the bottleneck part. Third, the batch normalization is replaced with the network deconvolution~\cite{Ye2020}. 
% In each dense block, the dilation factor of the initial convolution is set to one, and it doubles every time the next convolution is applied, as applied in WaveNet~\cite{Aaron2016WN}. 
We call this architecture \emph{RD3Net}; the architecture is illustrated in Figure \ref{fig:netarch}.    

\begin{figure}[t]
    \centering
    \centerline{\includegraphics[width=0.6\linewidth]{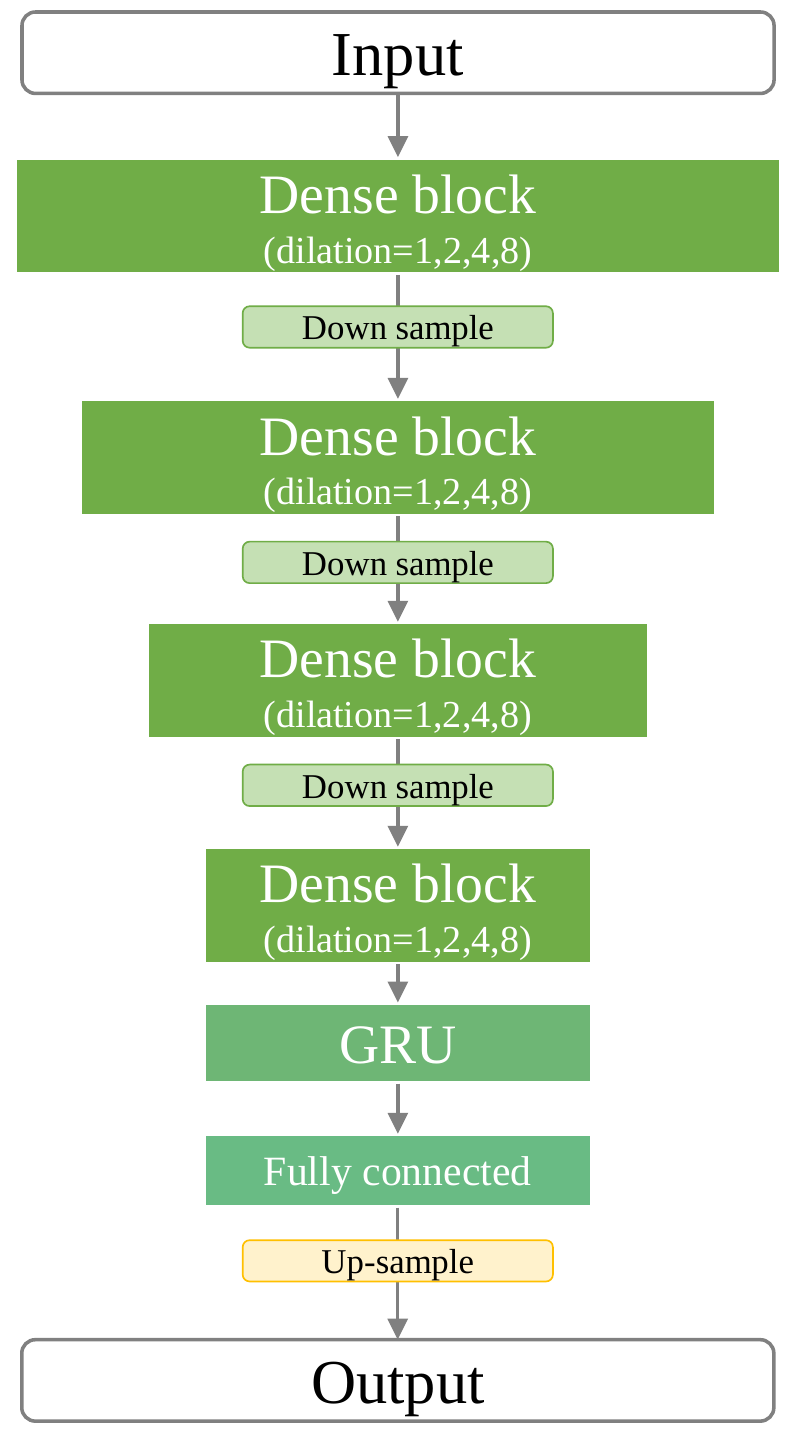}}
    \caption{Illustration of RD3Net architecture.}
    \label{fig:netarch}
\end{figure}

\subsection{Loss function}
\label{ssec:loss}
In the ACCDOA system, we solve the multi-output regression with a mean square error~(MSE) loss. In the other system using the two-stage training scheme, we use a binary cross entropy~(BCE) for the SED classification task and a masked MSE for the DOA regression task~\cite{cao2019polyphonic}. The latter is based on an MSE masked with the ground truth activations of each class, hence not contributing to the training when the corresponding sound event is not active.

\subsection{Post-processing}
\label{ssec:post}
% In the inference, the shift frame length is less than the input frame length, and the system outputs the average of each frame-wise output.
During the inference, we split the 60 seconds inputs into shorter segments with overlap, process each segment, and average the results of overlapped frames. 
To further improve the performance, we conduct a post-processing with the following procedure: rotating the FOA data, estimating the ACCDOA vectors, rotating the vectors back, and averaging the vectors of different rotation patterns. 
Similar to section~\ref{ssec:augmentation}, we consider eight rotations.

\subsection{Model ensemble}
\label{ssec:ensemble}

\begin{table}[t]
    \centering
    \caption{Ensemble configuration.}
    \vspace{1mm}
        \begin{tabular}{l|c|l}
        \toprule
        Name & \# of models & Base system \\
        \midrule
        \multirow{2}{*}{Ensemble 1} & \multirow{2}{*}{4} & ACCDOA w/ RD3Net $\times$3 \\
        & & Two-stage w/ RD3Net \\
        \midrule
        \multirow{2}{*}{Ensemble 2} & \multirow{2}{*}{5} & ACCDOA w/ RD3Net $\times$3 \\
        & & Two-stage w/ RD3Net, CRNN \\
        \midrule
        \multirow{2}{*}{Ensemble 3} & \multirow{2}{*}{5} & ACCDOA w/ RD3Net $\times$3 \\
        & & Two-stage w/ RD3Net  $\times$2 \\
        % Ensemble 2 & ACCDOA w/ RD3Net $\times$3, Two-stage w/ RD3Net, Two-stage w/ RCNN \\
        % Ensemble 3 & ACCDOA w/ RD3Net $\times$3, Two-stage w/ RD3Net  $\times$2 \\
        \bottomrule
        \end{tabular}
    \label{tb:ensemble}
\end{table}

\begin{table*}[t]
    \centering
    \caption{SELD performance of our systems evaluated using joint localization/detection metrics for the development set.}
    \vspace{1mm}
        \begin{tabular}{ll|cccc|cccc}
        \toprule
        & & \multicolumn{4}{c|}{Validation split} & \multicolumn{4}{c}{Testing split} \\
        Submission label & System & ${LE}_{CD}$ & ${LR}_{CD}$ & ${ER}_{20^{\circ}}$ & ${F}_{20^{\circ}}$ & ${LE}_{CD}$ & ${LR}_{CD}$ & ${ER}_{20^{\circ}}$ & ${F}_{20^{\circ}}$ \\
        \midrule
        - & Baseline FOA & $23.5^{\circ}$ & 62.0 & 0.72 & 37.7 & $22.8^{\circ}$ & 60.7 & 0.72 & 37.4 \\
        % Baseline MIC & $27.0^{\circ}$ & 62.6 & 0.74 & 34.2 & $27.3^{\circ}$ & 59.0 & 0.78 & 31.4 \\
        % \midrule
        - & ACCDOA w/ CRNN & $7.8^{\circ}$ & 79.9 & 0.34 & 77.2 & $8.9^{\circ}$ & 73.8 & 0.40 & 70.5 \\  % 4.21
        - & Two-stage w/ RD3Net & $8.6^{\circ}$ & 86.2 & 0.29 & 80.4 & $9.7^{\circ}$ & 78.2 & 0.38 & 73.0 \\  % 5.32&6.31
        Shimada\_SONY\_task3\_1 & ACCDOA w/ RD3Net & $7.0^{\circ}$ & 87.0 & 0.24 & 84.4 & $7.9^{\circ}$ & 80.5 & 0.32 & 76.8 \\  % 4.24
        \midrule
        Shimada\_SONY\_task3\_2 & Ensemble 1 & $6.3^{\circ}$ & 90.0 & 0.20 & 87.6 & $7.5^{\circ}$ & 82.9 & 0.29 & 79.4 \\  % Ens1.7.2
        Shimada\_SONY\_task3\_3 & Ensemble 2 & $6.3^{\circ}$ & 90.6 & 0.18 & 88.0 & $7.6^{\circ}$ & 83.7 & 0.28 & 79.9 \\  % Ens1.14
        Shimada\_SONY\_task3\_4 & Ensemble 3 & $6.4^{\circ}$ & 90.6 & 0.18 & 88.0 & $7.5^{\circ}$ & 83.5 & 0.29 & 80.0 \\  % Ens1.15
        \bottomrule
        \end{tabular}
    \label{tb:result}
\end{table*}

% An ensemble of the models of the two systems is used by averaging the outputs with weights. 
It is well known that averaging outputs of several models trained with different conditions such as initial parameters, stopping iteration, input features and model architectures often provides an improvement over the individual models.
Here, we perform the model ensemble by averaging the outputs with weights.
The weights are assigned to each class and model, thus the dimension of weights is $ N \times M$, where $N$ is the number of class and $M$ is the number of models. The weights are estimated by the stochastic gradient decent on the validation set using MSE criteria as the ACCDOA system.

The systems used for the ensemble is shown in Table \ref{tb:ensemble}. Here, CRNN means the convolutional
recurrent neural network architecture used in~\cite{cao2019polyphonic}.
Some of the models use PCEN~\cite{lostanlen2018per} with and without mel filter, cosIPDs, and sinIPDs~\cite{wang2018multi} as input features, instead of the amplitude spectrograms and IPDs.

\section{Experimental evaluation}
\label{sec:eval}
In this section, we show the experimental results of our systems on the development dataset.

\begin{table}[t]
    \centering
    \caption{SELD performance without and with polyphony for the development set.}
    \vspace{1mm}
        \begin{tabular}{l|cccc}
        \toprule
        & \multicolumn{4}{c}{Testing split} \\
        ACCDOA w/ RD3Net & ${LE}_{CD}$ & ${LR}_{CD}$ & ${ER}_{20^{\circ}}$ & ${F}_{20^{\circ}}$ \\
        \midrule
        Without polyphony & $6.7^{\circ}$ & 83.1 & 0.25 & 81.3 \\  % 4.24
        With polyphony & $8.6^{\circ}$ & 79.0 & 0.36 & 74.3 \\  % 4.24
        \bottomrule
        \end{tabular}
    \label{tb:result_polyphony}
\end{table}

\subsection{Experimental settings}
\label{ssec:setting}
We evaluated our approach on the development set of TAU Spatial Sound Events 2020 - Ambisonic using the suggested setup~\cite{politis2020dataset}. In the setup, four metrics were used for the evaluation~\cite{mesaros2019joint}. The first was the localization error ${LE}_{CD}$, which expresses the average angular distance between predictions and references of the same class. The second was a simple localization recall metric ${LR}_{CD}$ that expresses the true positive rate of how many of these localization estimates were detected in a class out of the total number of class instances. The next two metrics were the location-dependent error rate (${ER}_{20^{\circ}}$) and F-score (${F}_{20^{\circ}}$), where predictions were considered as true positives only when the distance from the reference is less than  $20^{\circ}$.
% considering that true positives were predicted only under a distance threshold of $20^{\circ}$ from the reference.

The sampling frequency was set to 24 kHz. The STFT was applied with a configuration of 20 ms frame length and 10 ms frame hop. The frame length of input to the networks was 1,024 frames. During the inference time, the frame shift length was set to 20 frames.
We used a batch size of 32. Each training sample was generated on-the-fly~\cite{erdogan2018investigations}. The learning rate was set to 0.001 and decayed 0.9 times every 20,000 iterations. We used Adam optimizer with a weight decay of~$10^{-6}$.

All final submitted systems were trained on the fold 3, 4, 5 and 6 of the dataset except the "Shimada\_SONY\_task3\_4" where one of the two-stage model in the ensemble was trained on the fold 1, 3, 4, 5 and 6. The the fold 2 was used for the validation set all the time.

\begin{figure}[t]
    \centering
    % \centerline{\includegraphics[width=0.98\linewidth]{./Figure/fold1_room1_mix006_ov1.pdf}}
    \centerline{\includegraphics[width=0.98\linewidth]{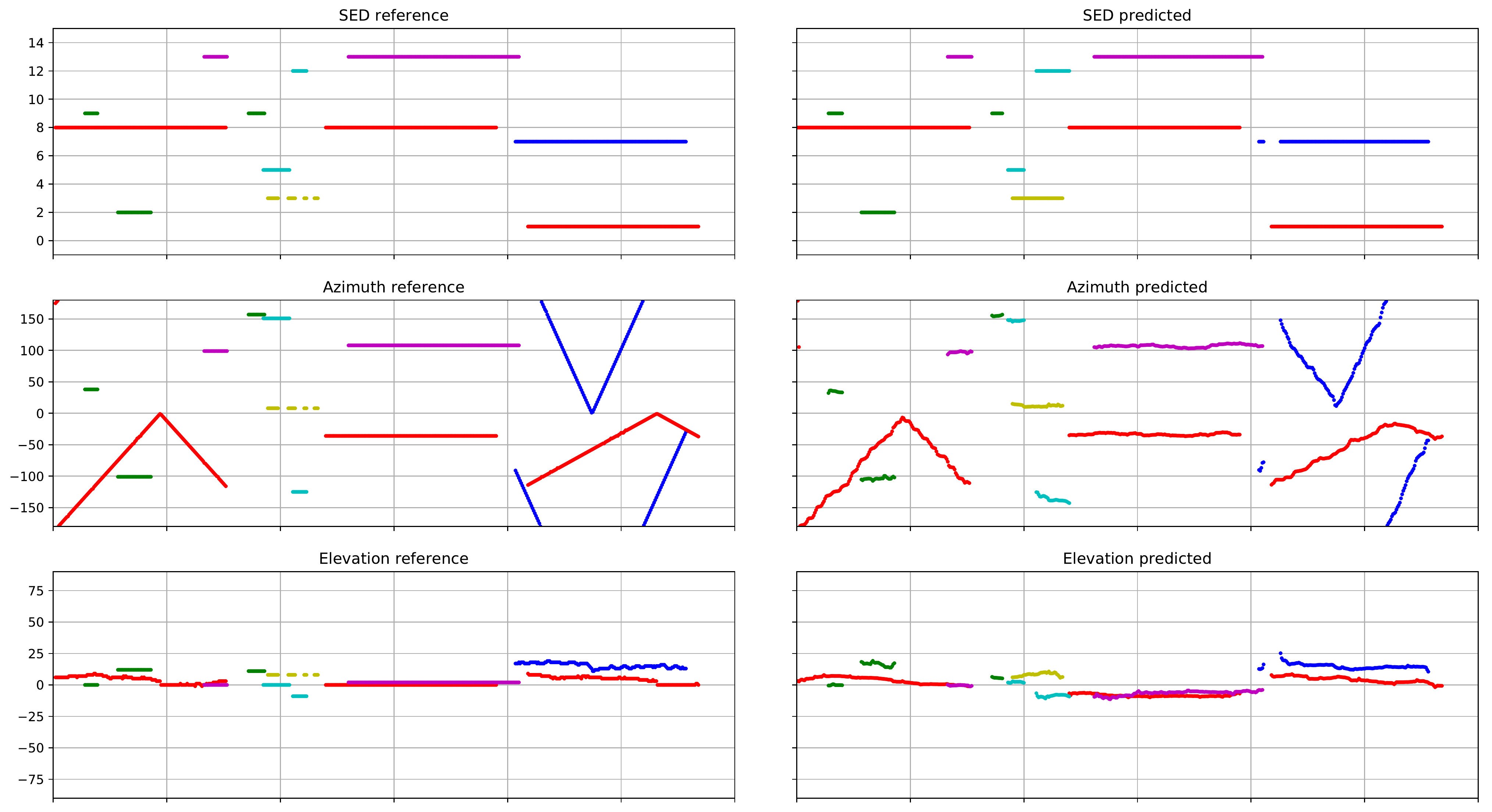}}
    \caption{Visualization of SELD output for ACCDOA w/ RD3Net.}
    \label{fig:visualize}
    \vspace{-2mm}
\end{figure}

\subsection{Experimental results}
\label{ssec:result}

Table~\ref{tb:result} shows the performance with the development set for our systems. As shown in the table, our systems outperformed the baseline for each metric by a large margin.
We compared the performances of RD3Net and CRNN used in~\cite{cao2019polyphonic} with the ACCDOA system. The results show significant improvements over CRNN in all metrics, demonstrating the advantage of RD3Net.
% RD3Net improved ${F}_{20^{\circ}}$ by 6.3 points from CRNN in the testing split.
We also compared the performances of the ACCDOA system and two-stage system. The ACCDOA system showed 2.3 points higher ${LR}_{CD}$ than the two-stage in the testing split, while the ACCDOA system improved ${F}_{20^{\circ}}$ by 3.8 points. This suggests that the the ACCDOA system is more effective in the location-aware detection.
% The unified framework make especially the location-aware detection metrics better.
Model ensemble improved ${F}_{20^{\circ}}$ by 3.2 points from the single model in the testing split.
Table~\ref{tb:result_polyphony} shows the performances of the ACCDOA system "Shimada\_SONY\_task3\_1" on recordings without and with polyphony. We observed that the performance on recordings without polyphony is better than with polyphony.

 An example of the proposed ACCDOA system output from the test split is visuarized in Fig.~\ref{fig:visualize}. Each event class is represented by a unique color. We can observe that our system performs joint detection, localization, and tracking of dynamic sources successfully in the recording.

% \subsection{Submission}
% \label{ssec:submit}
% The left side on Table~\ref{tb:result} shows the submission label.
% All systems were trained on the fold 3, 4, 5 and 6 from the dataset except the "Shimada\_SONY\_task3\_4" where one of the two-stage model in the ensemble was trained on the fold 1, 3, 4, 5 and 6. The the fold 2 is used for the validation set.

\section{Conclusion}
\label{sec:concl}
We presented our approach to DCASE2020 task 3, Sound Event Localization and Detection. Our systems use the ACCDOA representation to solve both SED and SEL tasks in a unified manner. Moreover, we proposed an efficient network architecture called RD3Net. Our systems showed superior performance over the baselines with a single model. Furthermore, we observed further improvement with an ensemble of the ACCDOA and two-stage systems. 

\section{Acknowledgement}
\label{sec:ack}
We would like to thank Yuichiro Koyama for the useful discussions on ACCDOA.

% -------------------------------------------------------------------------
% Either list references using the bibliography style file IEEEtran.bst
\bibliographystyle{IEEEtran}
\bibliography{refs}

\begin{thebibliography}{10}
\providecommand{\url}[1]{#1}
\def\UrlFont{\rmfamily}
\providecommand{\newblock}{\relax}
\providecommand{\bibinfo}[2]{#2}
\providecommand\BIBentrySTDinterwordspacing{\spaceskip=0pt\relax}
\providecommand\BIBentryALTinterwordstretchfactor{4}
\providecommand\BIBentryALTinterwordspacing{\spaceskip=\fontdimen2\font plus
\BIBentryALTinterwordstretchfactor\fontdimen3\font minus
  \fontdimen4\font\relax}
\providecommand\BIBforeignlanguage[2]{{%
\expandafter\ifx\csname l@#1\endcsname\relax
\typeout{** WARNING: IEEEtran.bst: No hyphenation pattern has been}%
\typeout{** loaded for the language `#1'. Using the pattern for}%
\typeout{** the default language instead.}%
\else
\language=\csname l@#1\endcsname
\fi
#2}}

\bibitem{cao2019polyphonic}
Y.~Cao, Q.~Kong, T.~Iqbal, F.~An, W.~Wang, and M.~D. Plumbley, ``Polyphonic
  sound event detection and localization using a two-stage strategy,''
  \emph{arXiv preprint arXiv:1905.00268}, 2019.

\bibitem{Takahashi16}
N.~Takahashi, M.~Gygli, B.~Pfister, and L.~V. Gool, ``Deep convolutional neural
  networks and data augmentation for acoustic event detection,'' in \emph{Proc.
  Interspeech}, 2016.

\bibitem{Takahashi2017AENet}
N.~Takahashi, M.~Gygli, and L.~{Van Gool}, ``Aenet: Learning deep audio
  features for video analysis,'' \emph{IEEE Trans. on Multimedia}, vol.~20, pp.
  513--524, 2017.

\bibitem{mazzon2019first}
L.~Mazzon, Y.~Koizumi, M.~Yasuda, and N.~Harada, ``First order ambisonics
  domain spatial augmentation for dnn-based direction of arrival estimation,''
  in \emph{Proc. of DCASE Workshop}, 2019.

\bibitem{park2019specaugment}
D.~S. Park, W.~Chan, Y.~Zhang, C.-C. Chiu, B.~Zoph, E.~D. Cubuk, and Q.~V. Le,
  ``{SpecAugment}: A simple data augmentation method for automatic speech
  recognition,'' \emph{Proc. of Interspeech}, pp. 2613--2617, 2019.

\bibitem{zhang2019data}
J.~Zhang, W.~Ding, and L.~He, ``Data augmentation and prior knowledge-based
  regularization for sound event localization and detection,'' in \emph{Tech.
  Report of DCASE Challenge}, 2019.

\bibitem{Takahashi20d3net}
\BIBentryALTinterwordspacing
N.~Takahashi and Y.~Mitsufuji, ``{D3Net: Densely connected multidilated
  DenseNet for music source separation},'' \emph{arXiv preprint
  arXiv:2010.01733}, 2020. [Online]. Available:
  \url{https://arxiv.org/abs/2010.01733}
\BIBentrySTDinterwordspacing

\bibitem{Ye2020}
C.~Ye, M.~Evanusa, H.~He, A.~Mitrokhin, T.~Goldstein, J.~A. Yorke,
  C.~Fermuller, and Y.~Aloimonos, ``Network deconvolution,'' in \emph{Proc.
  ICLR}, 2020.

\bibitem{lostanlen2018per}
V.~Lostanlen, J.~Salamon, M.~Cartwright, B.~McFee, A.~Farnsworth, S.~Kelling,
  and J.~P. Bello, ``Per-channel energy normalization: Why and how,''
  \emph{IEEE Signal Processing Letters}, vol.~26, no.~1, pp. 39--43, 2018.

\bibitem{wang2018multi}
Z.-Q. Wang, J.~Le~Roux, and J.~R. Hershey, ``Multi-channel deep clustering:
  Discriminative spectral and spatial embeddings for speaker-independent speech
  separation,'' in \emph{Proc. of IEEE ICASSP}, 2018, pp. 1--5.

\bibitem{politis2020dataset}
A.~Politis, S.~Adavanne, and T.~Virtanen, ``A dataset of reverberant spatial
  sound scenes with moving sources for sound event localization and
  detection,'' \emph{arXiv preprint arXiv:2006.01919}, 2020.

\bibitem{mesaros2019joint}
A.~Mesaros, S.~Adavanne, A.~Politis, T.~Heittola, and T.~Virtanen, ``Joint
  measurement of localization and detection of sound events,'' in \emph{Proc.
  of IEEE WASPAA}, 2019.

\bibitem{erdogan2018investigations}
H.~Erdogan and T.~Yoshioka, ``Investigations on data augmentation and loss
  functions for deep learning based speech-background separation,'' in
  \emph{Proc. of Interspeech}, 2018, pp. 3499--3503.

\end{thebibliography}

\end{sloppy}
\end{document}